\documentclass[aps,pre,showpacs,amsmath,amssymb,superscriptaddress,reprint,10pt]{revtex4-1}

\usepackage{amsthm}
\usepackage{bm}
\usepackage{verbatim}
\usepackage{graphicx}
\usepackage{url}
\usepackage[colorlinks=true,breaklinks=true,allcolors=blue]{hyperref}

\newcommand{\beq}{\begin{equation}}
\newcommand{\eeq}{\end{equation}}
\newcommand{\beqa}{\begin{eqnarray}}
\newcommand{\eeqa}{\end{eqnarray}}

\newcommand{\NN}{{\mathbb N}}
\newcommand{\RR}{{\mathbb R}}
\newcommand{\CC}{{\mathbb C}}
\newcommand{\EE}{{\mathbb E}}

\usepackage{xcolor}

\begin{document}

\title{Energy Landscape of the Finite-Size Mean-field 3-Spin Spherical Model}

\author{Dhagash Mehta}
\email{dbmehta@syr.edu}
\affiliation{Department of Physics, Syracuse University, Syracuse, NY 13244, USA}

\author{Daniel A.\ Stariolo}
\email{daniel.stariolo@ufrgs.br}
\affiliation{Instituto de F\'isica, Universidade Federal do Rio Grande do Sul and
National Institute of Science and Technology for Complex Systems, CP 15051, 91501-970 Porto Alegre, RS, Brasil.}

\author{Michael Kastner}
\email{kastner@sun.ac.za}
\affiliation{Institute of Theoretical Physics,  University of Stellenbosch, Stellenbosch 7600, South Africa}
\affiliation{National Institute for Theoretical Physics (NITheP), Stellenbosch 7600, South Africa}

\begin{abstract}
We study the $3$-spin spherical model with mean-field interactions and Gaussian random couplings. For moderate system sizes of up to 20 spins, we obtain all stationary points of the energy landscape by means of the numerical polynomial homotopy continuation method. On the basis of these stationary points, we analyze the complexity and other quantities related to the glass transition of the model and compare these finite-system quantities to their exact counterparts in the thermodynamic limit.
\end{abstract}

\maketitle

\section{Introduction}

In this work we analyze key characteristics of the energy landscape of the spherical $p$-spin glass model ($p$SM) \cite{crisanti.1992,KuPaVi1993} by means of the numerical homotopy continuation method. This method is capable of finding all stationary points of the energy function. However, since the number of stationary points grows exponentially with the system size $N$, one is restricted to rather small systems. The interest in the $p$SM resides in its connection with the theory of the glass transition. Although it is, strictly speaking, a spin glass model with disordered quenched couplings, its physics resembles in many respects that of structural glass formers and has been a key model in the study of the glass transition (for a review see \cite{CaCa2005}).

Its landscape properties have been extensively studied in the large-$N$ limit where, owing to its mean-field character, the model can be solved exactly. The number of stationary points of the so-called Thouless-Anderson-Palmer (TAP) free energy has been computed and characterized in Refs.\ ~\cite{KuPaVi1993,CrSo1995,CaGiPa1998,CrLeRi2003}. A key quantity is the complexity
\beq\label{complexity}
\Sigma(e)=\frac{1}{N}\ln\mathcal{N}(e)
\eeq
as a function of the energy density $e$, where $\mathcal{N}$ is the number of stationary points of the TAP free energy. Besides the number of stationary points, also their index $I$ turns out to be relevant, i.e., the number of negative eigenvalues of the Hesse matrix evaluated at a stationary point. For this reason it is convenient to also introduce the complexities
\beq
\Sigma_I(e)=\frac{1}{N}\ln\mathcal{N}_I(e),
\eeq
where $\mathcal{N}_I$ is the number of stationary points of the TAP free energy with index $I$. The picture that emerges is that the $p$SM has a high-energy regime characterized by the presence of exponentially many stationary points with large index. Asymptotically in the large-$N$ limit, the complexities $\Sigma_I$ are equal for all $I=0,\dotsc,N-1$. This regime corresponds to a high-temperature, or ``simple liquid'', regime. Below a certain threshold energy $e_\text{th}$ the number of minima becomes exponentially dominant over higher-index stationary points, and the system's dynamical behavior is governed by states near the bottom of the landscape \cite{Cavagna2001}. Interestingly, $e_\text{th}$ coincides with the energy at which a dynamic singularity is present, in the sense that relaxation times diverge in the large-$N$ limit. This dynamic singularity corresponds to the so-called mode coupling transition and is present in many glass forming models~\cite{BuCuKuMe1996,2002PhRvL..88e5502G,PhysRevLett.85.5356}. In the minima-dominated regime, for large $N$, the complexity of minima gives a finite contribution to the free energy of the system, called the configurational entropy.

In mean-field models like the one we are considering in the present article, divergent energy barriers lead to true ergodicity breaking at $e_\text{th}$. Although this is known not to be true anymore in more realistic models with short-range interactions, several short-range models still show signatures of a change of regimes in the energy landscape near the mode coupling transition. One such signature is the vanishing of the mean index density $\bar{i}$ near the mode coupling transition, where $\bar{i}(e)$ is obtained by averaging the index density $i=I/N$ of the Hesse matrix over all stationary points with energy $e$ \cite{2002PhRvL..88e5502G,PhysRevLett.85.5356}. In real systems it is expected that relaxation in the low energy regime will be driven by activation over barriers, due to the dominance of minima in the energy landscape probed by the system's dynamics. The complexity of minima stays positive in a restricted energy window below the threshold, diminishing continuously until a second characteristic energy below which the complexity changes sign. In the mean field picture this means that, for energy densities $e<e_\text{c}$, the number of minima becomes exponentially small (instead of exponentially large) for large $N$. Interestingly, this energy coincides with the point at which a replica calculation gives a replica symmetry breaking phase transition in the thermodynamic limit. At this point a true phase transition, hallmarked by a single step of replica symmetry breaking, occurs and the model enters a spin glass phase (or ``ideal glass phase'' in the structural glass terminology \cite{CrSo1995,DeSt2001}).

From a landscape perspective on thermodynamic properties of the model, the $p$SM has the striking feature that the equations defining the stationary points of the TAP free energy (which is a mean-field free energy function of local magnetizations) are formally identical to the equations determining the stationary points of the Hamiltonian \cite{KuPaVi1993,CaGiPa1998}. Therefore, in the thermodynamic limit, many temperature-dependent properties of the model can be inferred directly from an analysis of the energy landscape. In general (i.e., for models other than the $p$SM) such a duality of TAP free energy stationary points and Hamiltonian stationary points does not hold. The extrapolation of energy landscape characteristics (based on the Hamiltonian) to the thermal behavior of glassy systems (governed by the free energy) must be done with considerable care.

The energy landscape approach has greatly improved the understanding of the glass transition, and much of this success is due to the exact results for landscape-related quantities of the $p$SM in the thermodynamic limit. Nonetheless, many questions related to the way that barriers and timescales scale with size are relevant for real systems and are still poorly understood. The numerical homotopy continuation method, by virtue of its capacity of determining all stationary points of a complex polynomial, opens up a way of obtaining exact finite-system properties of the energy landscape of the $p$SM for system sizes up to $N=20$ (i.e., far from the thermodynamic limit). The main goals of the present study are (1) to characterize the landscape of a small cluster of spins, which can be considered as a prototype of a real, large-$N$ glass model, and approach in a systematic way the emergence of complexity as known in the large-$N$ case, and (2) to test how far the validity of the thermodynamic-limit results for the $p$SM extends to finite and even small system sizes.

\section{The mean-field $p$-spin Spherical Model}
The $p$-spin spherical model is defined by the Hamiltonian
\beq\label{phamilton}
H= -\frac{1}{p!}\sum_{i_1, i_2,\dots, i_p = 1}^{N} J_{i_1, i_2, \dots, i_p} \sigma_{i_1}\sigma_{i_2}\cdots \sigma_{i_p},
\eeq
subject to the spherical constraint
\beq\label{constraint}
\sum_{i=1}^{N}\sigma_{i}^2 = N
\eeq
that restricts the configurations $\sigma=(\sigma_1,\dotsc,\sigma_N)\in\RR^N$ to an $(N-1)$-sphere of radius $\sqrt{N}$ (hence the name of the model). The configuration space of the $p$SM is therefore effectively $(N-1)$-dimensional. The coupling constants $J_{i_1, \dotsc, i_p}$ are Gaussian random variables with zero mean and standard deviation $\sqrt{p!/2N^{p-1}}$. The Hamiltonian \eqref{phamilton} consists of $p$-spin interactions between all possible groupings of $p$ lattice sites. For such a fully-connected model the mean-field approximation is known to be exact, and for this reason the model itself is called the {\em mean-field} $p$SM. At least qualitatively, the phenomenology of the $p$SM is rather insensitive to the value of the integer parameter $p$, as long as $p\geq3$. For this reason we will restrict our numerical study to the case $p=3$ where the Hamiltonian of the 3-spin spherical model (3SM) is given by
\beq\label{3hamilton}
H= -\frac{1}{6}\sum_{i,j,k = 1}^{N} J_{i,j,k} \sigma_{i}\sigma_{j}\sigma_{k}.
\eeq

In order to compute stationary points of the Hamiltonian \eqref{phamilton} restricted to the spherical configuration space by the constraint \eqref{constraint}, it is convenient to resort to the formalism of Lagrange multipliers. We define the Lagrangian
\begin{equation}
L = H + \sigma_{0} \left(\sum_{i=1}^{N}\sigma_{i}^2 - N \right),
\end{equation}
where $\sigma_0$ is a Lagrange multiplier. The stationary points of the constrained system are then the solutions of the system of equations
\begin{equation}\label{stat}
\frac{\partial L}{\partial \sigma_{i}} = 0, \hspace{2cm} i=0,\dotsc, N.
\end{equation}
For the 3SM these equations are
\begin{equation}\label{stationaryeqs}
\sum_{j,k = 1}^N J_{i,j,k} \sigma_j\sigma_k + 6e\sigma_i=0,
\end{equation}
where $e=H/N$ is the energy density at a stationary point \cite{CaGiPa1998}. The stability of a stationary point is governed by the Hesse matrix
\beq\label{Hessian}
\tilde{\mathcal{H}}_{lm} \equiv \frac{\partial^2 L}{\partial \sigma_l \partial \sigma_m} = -\sum_{k=1}^N J_{k,l,m} \sigma_k- p\,e\, \delta_{l,m}
 \eeq
with $l,m=1,\dotsc,N$, where $\delta_{l,m}$ denotes Kronecker's symbol. The matrix $\tilde{\mathcal{H}}$ as defined in \eqref{Hessian} is an $N \times N$ matrix. Evidently, this cannot be the correct Hesse matrix of the 3SM: as mentioned earlier, the spherical constraint \eqref{constraint} restricts the configuration space of the model to an $(N-1)$-dimensional sphere. Hence the correct Hessian $\mathcal{H}$, i.e., the one constrained to the spherical configuration space, should be an $(N-1) \times (N-1)$ matrix. This can be accounted for by eliminating from $\tilde{\mathcal{H}}$ the eigendirection perpendicular to the surface of the configuration space manifold. For a spherical configuration space such a perpendicular eigenvector is radial for any given stationary point. It is possible to prove that the eigenvalue corresponding to this radial eigenvector is equal to $3e$ for the 3SM \cite{CrLeRi2003}. Knowledge of this radial eigenvalue allows us to easily obtain the index of the correct (constrained) Hessian $\mathcal{H}$ from the unconstrained one,
\beq
I(\mathcal{H}(\sigma^\text{s}))=I(\tilde{\mathcal{H}}(\sigma^\text{s}))-\Theta(-e),
\eeq
where $\sigma^\text{s}=(\sigma_1^\text{s},\dotsc,\sigma_N^\text{s})$ is a solution of \eqref{stationaryeqs} and $e=H(\sigma^\text{s})/N$ is the corresponding energy density. Similarly, the determinant of the constrained Hessian can be obtained from $\tilde{\mathcal{H}}$,
\beq
\det(\mathcal{H})=\frac{\det(\tilde{\mathcal{H}})}{3e}\qquad\text{for $e\neq0$}.
\eeq

\section{Numerical Polynomial Homotopy Continuation Method}
\label{sec_homotopy}
The stationary points of the 3SM are the solutions of the $N$ coupled nonlinear equations in \eqref{stationaryeqs}. Since an analytic solution of these equations is unlikely to be feasible, we resort to a numerical technique. Due to the polynomial character of \eqref{stationaryeqs}, the so-called numerical polynomial homotopy continuation (NPHC) method \cite{SW:95} is particularly suitable. This method has the virtue of being able to find {\em all}\/ the solutions of a given system of polynomial equations, and has been applied in the past to a variety of problems in particle theory and statistical mechanics \cite{Mehta:2009,Mehta:2009zv,Mehta:2011xs,Mehta:2011wj,Kastner:2011zz,Casetti-Mehta:2012,2012PhRvE..85f1103M,Maniatis:2012ex,Mehta:2012wk,Hughes:2012hg,MartinezPedrera:2012rs,He:2013yk}. A drawback of the NPHC method is that it is restricted to fairly small system sizes. A detailed description of the method can be found in \cite{Mehta:2011xs} or in Sec.\ III of \cite{2012PhRvE..85f1103M}.

Running the NPHC method on a computer, we consider a configuration $(\sigma_1,\dotsc,\sigma_N)$ to be a numerical solution of \eqref{stationaryeqs} if it satisfies these equations with tolerance not exceeding $10^{-10}$. The NPHC method searches solutions in $\CC^N$ and, since we are interested only in real solutions, we have to dispose of all the complex ones as a last step. We consider a solution to be real if the absolute value of the imaginary part of each component $\sigma_i$ does not exceed the tolerance $10^{-6}$. We have checked that the choice of this tolerance does not affect the number of real solutions found, i.e., this number is robust for the problem at hand. All stationary points can be further refined to arbitrary numerical precision if necessary. Moreover, we have used an efficient implementation of Smale's $\alpha$-theory \cite{2010arXiv1011.1091H} to certify that each of the numerical solutions is indeed in the quadratic convergence of the associated actual solution (see \cite{Mehta:2013zia} for an introduction of the method from the potential energy landscape point of view).

\section{Results}
\label{s:results}

By means of the NPHC method we computed all stationary points of the 3SM for system sizes $N$ between 14 and 20 and, for each value of $N$, ten different realizations of the random couplings were used. In the Introduction we had outlined the relevance of the stationary points and their indices for the relaxational properties of glassy systems. In the following sections we will analyze for the finite-size 3SM the most relevant properties, including complexities, mean indices, and the Hesse determinant, and compare them to analytic predictions for their infinite-system counterparts.

\subsection{Number of stationary points}
\begin{figure}\centering
\includegraphics[width=0.7\linewidth]{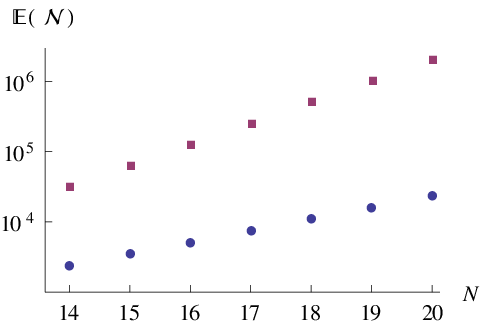}
\includegraphics[width=0.7\linewidth]{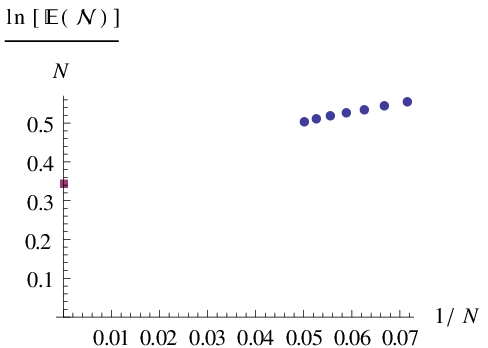}
\caption{(color online) Top: Disorder average $\EE(\mathcal{N})$ of the total number $\mathcal{N}$ of complex (red squares) and real (blue dots) stationary points on a logarithmic scale, plotted {\em vs}.\ the system size $N$. Both numbers grow exponentially with $N$. Bottom: Logarithmic density of $\EE(\mathcal{N})$ for the real stationary points, plotted {\em vs}. the inverse system size (blue dots). The analytic result in the thermodynamic limit is indicated by a red square.}
\label{totnumsols}
\end{figure}

In Fig.\ \ref{totnumsols} (top) we show the total numbers of complex and real stationary points, respectively, averaged over the disorder samples. The linear increase in this logarithmic plot nicely illustrates that the total number of stationary points grows exponentially with $N$. While this is true in many models, the relevance of this behavior in the 3SM model is that not only the energy function, but the free energy itself has the same structure of stationary points. This is in stark contrast with the majority of statistical systems for which the number of stationary points of the free energy in order parameter space is usually much smaller. This is intimately connected to the glassy behavior of the 3SM at low temperatures/energies. The large number of complex stationary points is the limiting factor for applying the NPHC method to larger system sizes.

In the large-$N$ limit, an analytic expression is known for the disorder average  $\EE(\mathcal{N})$ of the total number $\mathcal{N}$ of real solutions,
\beq
\lim_{N\to\infty}\frac{1}{N}\ln\EE(\mathcal{N})=\frac{1}{2}\ln2,
\eeq
(Eq.\ (2.20) of Ref.\ \cite{AuArCe2011}). In Fig.\ \ref{totnumsols} (bottom) we plot $\ln\EE(\mathcal{N})/N$ {\em vs}.\ $1/N$ for our finite-system data. Although not an exact match, the numerical data tend nicely towards the exact limiting value, even for the very small system sizes considered.

\subsection{Complexity {\em vs}.\ energy}

Resolving the number of stationary points with respect to the energy, we obtain the complexity $\Sigma(e)$ as introduced in \eqref{complexity}. Although Eq.\ \eqref{complexity} captures the essential idea of the complexity, a technical difficulty is hidden in the number $\mathcal{N}(e)$ of stationary points at energy density $e$: Rigorously speaking, $\mathcal{N}(e)$ will be zero almost everywhere. To obtain a well-defined quantity, some kind of binning would be necessary, for example by defining $\mathcal{N}(e)$ to be the number of stationary points in the interval $[e,e+\Delta]$ for some small $\Delta>0$. A drawback of this approach is that an arbitrary parameter $\Delta$ is introduced, and the effect of the choice of $\Delta$ is difficult to control.

\begin{figure}\centering
\includegraphics[width=0.7\linewidth]{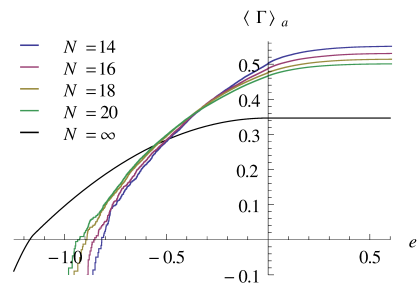}
\caption{(color online) The annealed average of the cumulative complexity $\Gamma$ plotted {\em vs.}\ the energy density $e$ for various system sizes $N$. For comparison, the analytic large-$N$ result is shown as a black line.}
\label{GammaAll}
\end{figure}

This problem can be circumvented by resorting to the cumulative complexity
\beq\label{cumcom}
\Gamma(e)=\frac{1}{N}\ln\mathcal{M}(e),
\eeq
where
\beq
\mathcal{M}(e)=\left|\left\{\sigma\,\bigg|\,\frac{\partial H}{\partial \sigma_i}(\sigma)=0\;\forall i,\;\;\frac{H(\sigma)}{N}\leq e\right\}\right|
\eeq
is the number of stationary points with energies not larger than $e$. The cumulative complexity $\Gamma(e)$, though discontinuous, is a well-defined, monotonically increasing function for all finite system sizes $N$. Also in analytical calculations the cumulative complexity $\Gamma$ is usually preferred to its non-cumulative counterpart $\Sigma$, and exact analytic expressions are known for the $p$SM in the thermodynamic limit (Eq.\ (2.15) of \cite{AuArCe2011}). The cumulative complexity is a self-averaging quantity in the large-$N$ limit, but since our system sizes are far from that limit, we calculated {\em quenched averages}
\begin{equation}
\langle\Gamma\rangle_\text{q}=\frac{1}{N}\mathbb{E}(\ln\mathcal{M})
\end{equation}
as well as {\em annealed averages}
\begin{equation}\label{e:annealed}
\langle\Gamma\rangle_\text{a}=\frac{1}{N}\ln\mathbb{E}(\mathcal{M}),
\end{equation}
where $\mathbb{E}$ denotes averaging over 10 disorder realizations \cite{CaGiPa1998}. In Fig.\ \ref{GammaAll} we show $\langle\Gamma\rangle_\text{a}$ for various system sizes $N$, together with the thermodynamic limit result. Although, for the small $N$ considered, the finite-system data are not close to the thermodynamic limit result for most $e$, the trend we observe is correct: For energies $e$ less than about $-0.55$ the finite-system data lie below the infinite system result, but become larger as $N$ grows; for $e$ larger than about $-0.45$ the finite-system data lie above the infinite system result, but decrease with increasing $N$. The quenched average $\langle\Gamma\rangle_\text{q}$ (not shown) turns out to be very similar to $\langle\Gamma\rangle_\text{a}$ and the two are almost indistinguishable on the scale of Fig.\ \ref{GammaAll} for most $e$. The only visible difference is that, for energies smaller than the ground state of any one of the disorder realizations, $\langle\Gamma\rangle_\text{q}$ is ill-defined (or $-\infty$). As a consequence, the quenched average $\langle\Gamma\rangle_\text{q}$ is nonnegative, and convergence to the negative values of the analytic result in the thermodynamic limit cannot occur. For this reason the annealed average $\langle\Gamma\rangle_\text{a}$ appears more suitable for such a comparison.

\begin{figure}\centering
\includegraphics[width=0.7\linewidth]{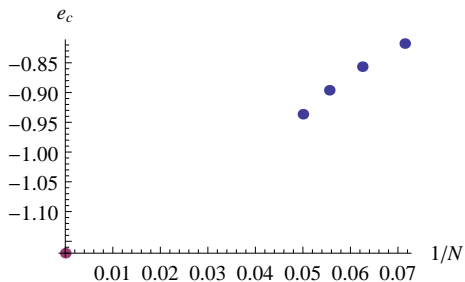}
\caption{(color online) Finite-system critical energies $e_\text{c}$, determined as the energy values at which the annealed average $\langle\Gamma\rangle_\text{a}$ in Fig.\ \ref{GammaAll} changes sign. For comparison, the thermodynamic limit value $e_\text{c} \approx -1.17$ is plotted in red.}
\label{ec_vs_N}
\end{figure}

In particular, the critical energy $e_\text{c}$, i.e. the energy at which the complexity goes to zero in the thermodynamic limit,  can be defined approximately, for finite $N$, as the point at which $\langle\Gamma\rangle_\text{a}$ changes sign. This point can be easily read off, even for the small system sizes studied here for which there is no rigorous relation with a true phase transition. As mentioned in the Introduction, the infinite-system value of $e_\text{c}$ marks the transition to a spin glass phase. For the $p$SM in the thermodynamic limit, an implicit analytic expression for the critical energy is known \cite{AuArCe2011}, yielding the approximate numerical value $e_\text{c} \approx -1.17$. The finite-system critical energies are plotted {\em vs.}\ the inverse system size $1/N$ in Fig.\ \ref{ec_vs_N}. Despite the small system sizes, the results indicate a trend which appears to be nicely consistent with the analytic thermodynamic limit value. It is important to note that the finite $N$ critical energies previously defined do not coincide with the
ground state energy of the clusters.

\subsection{Index-resolved number of stationary points}
\label{s:I-resolvedStatPoints}

The index $I$ of a stationary point $\sigma^\text{s}$, i.e., the number of negative eigenvalues of the Hesse matrix $\mathcal{H}(\sigma^\text{s})$, is known to
be a relevant quantity for the physical properties of the system under consideration. In particular, minima are stationary points with $I=0$ and correspond to regions in phase space in which a trajectory may be trapped, which in turn is relevant for glassy dynamics. Higher-index stationary points, in contrast, allow the system to diffuse more easily due to their inherent instability.

\begin{figure}\centering
\includegraphics[width=0.7\linewidth]{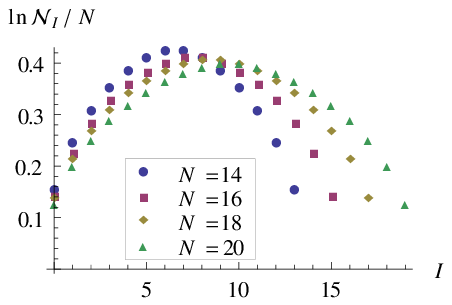}
\includegraphics[width=0.7\linewidth]{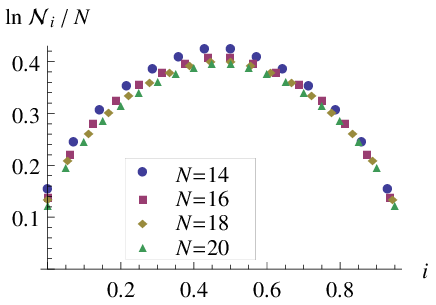}
\caption{(color online) The total number of stationary points {\em vs.}\ index $I$ (top) and {\em vs.}\ index density $i$ (bottom) for various system sizes $N$.}
\label{numsolsind}
\end{figure}

In Fig.\ \ref{numsolsind} (top) we plotted, for several system sizes $N$, the (logarithm of) total number of stationary points {\em vs.}\ the index $I$. The number of stationary points has a maximum at and is symmetric around $I=N/2$. This symmetry is a direct consequence of the antisymmetry of the Hamiltonian \eqref{3hamilton} of the 3SM: For every stationary point $\sigma^\text{s}= (\sigma_0^\text{s},\sigma_1^\text{s},\dotsc,\sigma_N^\text{s})$ with index $I(\sigma^\text{s})$, the point $\tilde{\sigma}^\text{s}= (-\sigma_0^\text{s},-\sigma_1^\text{s},\dotsc,-\sigma_N^\text{s})$ is also stationary and its index is $I(\tilde{\sigma}^\text{s})=N-I(\sigma^\text{s})$.

For the $p$SM in the thermodynamic limit, analytic results are available for the mean total number $\mathcal{N}_I$ of stationary points of a given index $I$,
\begin{equation}\label{e:N_I}
\lim_{N\to\infty}\frac{1}{N}\ln\EE(\mathcal{N}_I) = \frac{1}{2}\ln2-\frac{1}{3}\approx0.013,
\end{equation}
see Eq.\ (2.21) of \cite{AuArCe2011}. Interestingly, this result is independent of the index $I$. The numerical values for $\mathcal{N}_I$ from our finite-system data are much larger than the infinite-system results, but the scaling behavior with $N$ is roughly consistent with the expectation for small indices $I$.
 For $I\geq N/2$, the curves in Fig.\ \ref{numsolsind} (top) bend downwards as a consequence of the earlier mentioned symmetry of the 3SM Hamiltonian \eqref{3hamilton}. This strong finite-size effect can be reduced by considering the mean total number $\mathcal{N}_i$ of stationary points of a given index density $i$ [see Fig.\ \ref{numsolsind} (bottom)]. The trend of the data in Fig.\ \ref{numsolsind} (bottom), indicates a decrease of $\mathcal{N}_i$ with increasing system size, but our data do not allow an extrapolation to the thermodynamic limit.

Similar to Fig.\ \ref{numsolsind} (bottom), a unimodal shape of $\ln\mathcal{N}_i/N$ {\em vs.}\ $i$ has previously been observed for several other models. Examples include the $XY$ model with power law interactions for up to 13 degrees of freedom \cite{Mehta:2011xs}; the one-dimensional $XY$ model, for which all the stationary points were found analytically for any $N$ \cite{Mehta:2010pe, vonSmekal:2007ns}; the random-phase $XY$ model in two dimensions, for which all stationary points were computed for small square lattices (up to a $3\times 3$) \cite{ Mehta:2009zv, Hughes:2012hg}; Lennard-Jones clusters for up to 14 atoms \cite{2002JChPh.116.3777D}. In Ref.\ \cite{2003JChPh.11912409W}, for large bulk systems consisting of weakly interacting sub-systems, $\mathcal{N}_I$ was analytically shown to follow a binomial distribution in $I$.

\subsection{Index-resolved complexity}

\begin{figure}\centering
\includegraphics[width=0.7\linewidth]{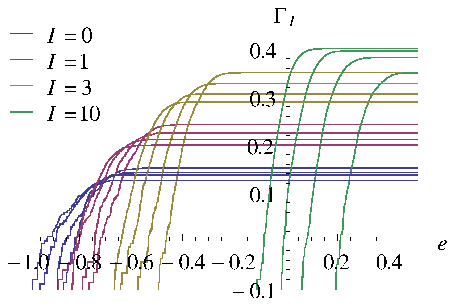}
\includegraphics[width=0.8\linewidth]{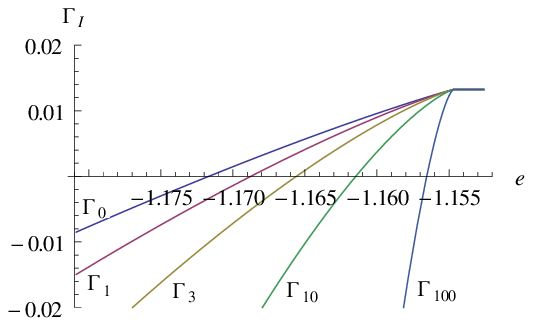}
\caption{(color online) Top: The cumulative complexities $\Gamma_I$ of stationary points of index $I=0$, 1, 3, and 10 as functions of the energy density $e$ for system sizes $N=14$, 16, 18, and 20. Among the curves of a given index $I$, the energy $e_\text{c}^I$ at which $\Gamma_I$ changes sign decreases with increasing system size, and this regularity allows one to easily relate system sizes to the various curves in the plot. Bottom: The corresponding thermodynamic limit results.}
\label{GammaI_vs_e}
\end{figure}

In Fig.\ \ref{GammaI_vs_e} (top) the cumulative complexities $\Gamma_I$ of stationary points of index $I=0$, 1, 3, and 10 are shown as functions of the energy density $e$ for various system sizes. Not unexpectedly, lower-index stationary points are more numerous at low energies, higher-index stationary points at high energies. Below a certain threshold value around $e=-0.9$, only minima (stationary points of index $I=0$) are present.

Also for the index-resolved complexities $\Gamma_I$ analytic results are known in the thermodynamic limit, see e.g. Eq.\ (2.16) of \cite{AuArCe2011}. For a convenient comparison we show a plot of these results in Fig.\ \ref{GammaI_vs_e} (bottom). For energies $e>e_\text{th}=-2/\sqrt{3}\approx-1.155$ above the threshold energy, the cumulative complexities $\Gamma_I$ in the thermodynamic limit are constant in $e$ and independent of $I$,
\begin{equation}
\Gamma_I(e)=\ln2/2-1/3\quad\text{for all $I$ and $e\geq e_\text{th}$}.
\end{equation}
Below $e_\text{th}$ the inequality $\Gamma_{I-1}(e)>\Gamma_I(e)$ holds for all $I=1,\dotsc,N$. This inequality implies that, in the thermodynamic limit, the complexity $\Gamma_0(e)$ of minima dominates the total complexity $\Gamma(e)$ for all $e<e_\text{th}$. Due to this dominance of minima, true ergodicity breaking occurs below this threshold, and the diffusivity of the systems goes to zero. This makes the threshold energy $e_\text{th}$ a relevant characteristic of the dynamical behavior of the $p$SM \cite{KuPaVi1993,CuKu1993,CrSo1995}.

Strictly speaking, the threshold energy $e_\text{th}$ is defined only in the thermodynamic limit where the heights of the barriers between stationary points diverge. For finite $N$ the barrier height remains finite, ergodicity is restored, and therefore one cannot expect a true diffusion arrest in the dynamics to take place at a particular energy. Nonetheless it would be interesting to study a finite-system analog of the threshold energy level in the $p$SM. For some models which in certain respects behave similar to the $p$SM, such a finite-system analog of $e_\text{th}$ can be defined as the energy below which a crossover takes place towards extremely slow dynamics. This slowing down is attributed to the suppression of escape directions in the energy landscape \cite{2002PhRvL..88e5502G,PhysRevLett.85.5356}. For the rather small system sizes at our disposal, we expect the relaxational dynamics of the system to be ruled by a spectrum of relaxation time scales that are related to the $N$-dependent heights of energy barriers in the low-energy regime dominated by minima. It would be interesting to analyze the dynamics of small clusters of spins in this regime and try to interpret the results in the light of the energy landscape properties described here. This could shed light on the mechanism that leads, in the large-system limit, to the characteristic two-step relaxation of glassy systems.

For the small system sizes we have studied, the finite-$N$ results do not even qualitatively resemble the thermodynamic limit results. For small indices $I\ll N$, we at least observe the correct trend, as the large-$e$ value of $\Gamma_I$ is decreasing with increasing system size $N$. The trend is reversed (and inconsistent with the large-$N$ result) for larger values of $I$, but this finite-size behavior can be attributed to similar reasons as discussed at the end of Sec.\ \ref{s:I-resolvedStatPoints}.

\subsection{Cumulative mean index density}

\begin{figure}\centering
\includegraphics[width=0.7\linewidth]{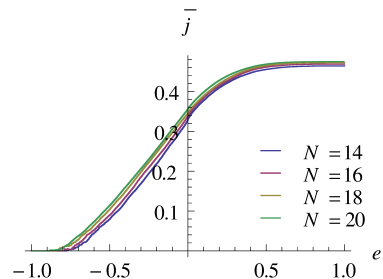}
\caption{(color online) Cumulative mean index density $\bar{j}$ {\em vs.}\ energy density $e$ for various system sizes $N$.}
\label{mean-index}
\end{figure}

One indication of the occurrence of a mode coupling transition at some value of the energy density $e$ is, as mentioned in the Introduction, the vanishing of the mean index density $\bar{i}(e)$ \cite{KuLa1996,2002PhRvL..88e5502G,PhysRevLett.85.5356,2002JChPh.116.3777D}, as obtained by averaging the index density $i=I/N$ of the Hesse matrix over all stationary points with energy $e$. For the reasons explained above, we prefer to use a cumulative version of this quantity,
\begin{equation}
\bar{j}(e) = \sum_{\{\sigma^\text{s}:\,H(\sigma^\text{s})\leq Ne\}}\frac{I(\sigma^\text{s})}{N},
\end{equation}
defined as the index density averaged over all stationary points with energy densities not larger than $e$. Results for $\bar{j}$, averaged over 10 disorder realizations are shown for system sizes $N=14,\dotsc,20$ in Fig.\ \ref{mean-index}. The cumulative mean index density is found to be a monotonically increasing function of the energy density $e$, saturating at a value of $1/2$ with increasing energy. This limiting value is again a consequence of the symmetries of the 3SM. Within the narrow range of system sizes we were able to probe, the $N$-dependence is very weak.

\begin{figure}\centering
\includegraphics[width=0.7\linewidth]{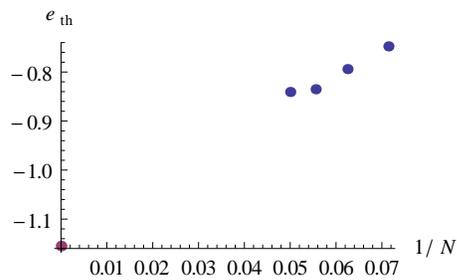}
\caption{(color online) Finite-system threshold energies $e_\text{th}$ as obtained from fits to the cumulative mean index densities $\bar{j}(e)$ in Fig.\ \ref{mean-index}. An extrapolation to small values of $1/N$ appears at least not inconsistent with the thermodynamic limit value $e_\text{th}=-2/\sqrt{3}\approx-1.155$.}
\label{eth_vs_N}
\end{figure}

As previously remarked, in the large-$N$ picture, the threshold energy $e_\text{th}$ below which the landscape is dominated by minima plays an important role. Naively one might try to identify the finite-system counterpart of $e_\text{th}$ with the value of $e$ at which $\bar{j}(e)$ starts deviating from zero. 
Here we follow a different approach: In the large-$N$ limit and asymptotically for $e\gtrsim e_\text{th}$, the mean index density $\bar{i}(e)$ is expected to grow like a power law of the distance from the threshold energy (Eq.\ (55) of \cite{KuLa1996}). Accordingly, a similar power law holds for the cumulative mean index density,
\beq\label{baripowerlaw}
\bar{j}(e) \propto (e-e_\text{th})^\alpha\qquad\text{for $e\geq e_\text{th}$}
\eeq
with exponent $\alpha=3/2$. Fitting such a power law to the data of Fig.\ \ref{mean-index}, we find that the optimal fit parameters $\alpha\approx1.3$ differs somewhat from the expected thermodynamic limit value. The finite-system threshold energies $e_\text{th}$, though not an exact match, are at least in reasonable agreement with the large-$N$ predictions. In Fig.\ \ref{eth_vs_N}, $e_\text{th}$ is plotted {\em vs.}\ inverse system size, and we observe a correct trend towards the infinite-system value $e_\text{th}=-2/\sqrt{3}\approx-1.155$.



\subsection{Hesse determinant}

In order to further characterize the stationary points of the Hamiltonian \eqref{3hamilton}, we computed the determinant of the Hesse matrix at each of the stationary points. Similar, but complementary, to the index of a stationary point, also the Hesse determinant represents a way of projecting the information contained in the Hesse matrix onto a single number. In energy landscape studies, the Hesse determinant has been used previously in the study of phase transitions, also in the absence of disorder. One of the key results in this context relates the vanishing of the Hesse determinant in the thermodynamic limit to the occurrence of a phase transition \cite{KaSchne08,KaSchneSchrei08}. More precisely, under some technical assumption, in these references it was shown that, if there exists a sequence of stationary points $\sigma^\text{s}_{(N)}$ for $N\in\NN$ such that the energy density of this sequence converges,
\beq
e_\text{c}=\lim_{N\to\infty}H(\sigma^\text{s}_{(N)})/N,
\eeq
and such that
\beq\label{KSS}
\lim_{N\to\infty}\bigl|\det\mathcal{H}(\sigma^\text{s}_{(N)})\bigr|^{1/N}=0,
\eeq
then the stationary points may induce a phase transition at the energy density $e_\text{c}$ in the thermodynamic limit. Eq. \ref{KSS} implies that stationary points have to become sufficiently ``flat'' in the thermodynamic limit in order to be capable of inducing a phase transition.
This scenario was confirmed in a number of exactly solvable long-range interacting models (mostly with mean-field interactions) \cite{KaSchne08,KaSchneSchrei08,PhysRevE.83.031114}, and it also led to the analytic prediction of the exact transition energy of the self-gravitating ring model \cite{PhysRevE.80.060103}.

\begin{figure}\centering
\includegraphics[width=0.7\linewidth]{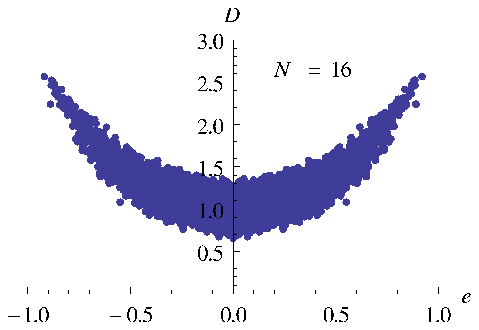}
\includegraphics[width=0.7\linewidth]{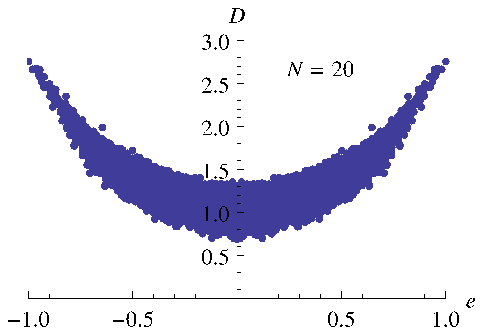}
\caption{(color online) For all stationary points $\sigma^\text{s}$ of our 10 disorder realizations of the 3SM, the pairs $(e,D)$ are plotted, where $e=H(\sigma^\text{s})/N$ is the energy density and $D=\bigl|\det\mathcal{H}(\sigma^\text{s})\bigr|^{1/N}$ is the rescaled Hesse determinant at a stationary point. The top plot is for system size $N=16$, the bottom one for $N=20$.}
\label{dets}
\end{figure}

On the basis of finite-system data, it is of course not possible to construct infinite sequences of stationary points $\sigma^\text{s}_{(N)}$ for $N\in\NN$. But we can try and see whether, for the small system sizes at our disposal, the Hesse determinant exhibits a tendency to vanish at some value of the energy, possibly close to the threshold energy or critical energy of the 3SM. To investigate this, we have computed Hesse determinants and energy densities for all stationary points of all our 10 disorder realizations of the model. The data, plotted in Fig.\ \ref{dets}, do not reveal any indication of a vanishing Hesse determinant. Moreover, comparing the data for system sizes $N=16$ (top) and $N=20$ (bottom) in Fig.\ \ref{dets}, no significant size-dependence is visible.

This is a negative result, and its interpretation is not entirely clear. On the one hand, the absence of any significant differences between the plots for $N=16$ and $N=20$ may be seen as an indication that the range of system sizes we are able to deal with is way too small to observe finite-size scaling effects that allow for an extrapolation towards the thermodynamic limit. But it may of course also be that, even for much larger systems, the Hesse determinant remains bounded away from zero. Note that such a scenario would not be in contradiction with the criterion \eqref{KSS}: This criterion specifies necessary conditions for a sequence of stationary points to induce a phase transition. And although examples are known where such stationary points indeed are responsible for the occurrence of a phase transition, this need not be the case for other models. And indeed, several model systems are known in which stationary points are not related to the occurrence of a phase transition \cite{BaroniMSc,PhysRevE.70.036125,PhysRevE.70.041101,PhysRevLett.93.150601,PhysRevE.72.016122,PhysRevE.72.056134,PhysRevE.76.051119,Kastner:2011zz}, and this might also be the case for the 3SM.

In summary, the exponentially (in $N$) large number of stationary points renders the criterion \eqref{KSS} very impractical for a naive numerical approach like ours. Only with intuition about the relevant types of stationary points, and a numerical algorithm focusing exclusively on those points, might a numerical test of the condition \eqref{KSS} be viable.

\section{Discussion and Conclusions}

We have studied the potential energy landscape of the mean-field $p$-spin spherical model ($p$SM) for $p=3$. By means of the numerical polynomial homotopy continuation method, we obtained a complete characterization of the energy landscape, in the sense of computing all stationary points of the Hamiltonian \eqref{3hamilton} for a number of disorder realizations. Since the number of stationary points grows exponentially with the system size $N$, the capacity to compute all the stationary points is simultaneously a virtue and a limiting factor, as it restricts the method to rather small system sizes (up to $N=20$ in our study).

The $p$SM is known as a prototype of a fragile glass forming system and for several of its landscape characteristics, like the number of stationary points, indices, and associated complexities, analytical results are known in the thermodynamic limit. The research reported in the present article complements these thermodynamic limit results with their finite-system counterparts. The aim of such a study is to gain insight into how certain infinite-system properties of the model emerge from finite $N$ clusters. In particular, when dealing with mean field models which show artificially divergent energy barriers when $N\to\infty$, the finite-size scaling of the barrier height with $N$ is a key quantity for understanding the slow dynamics of the model. For the small system sizes studied, we found results that, though roughly compatible with large-$N$ results, show significant finite-size effects.

The index-resolved number of stationary points, $\mathcal{N}_I$, shows a maximum at $I=N/2$ and, due to a symmetry of the Hamiltonian \eqref{3hamilton}, is symmetric around this point. At variance with this result, analytical calculations in the thermodynamic limit give just a constant value, independent of $I$. While the change with increasing $N$ is roughly compatible with the asymptotic result, we can not draw a strong conclusion about this point which deserves further study with larger systems.

When studying the cumulative complexities $\Gamma(e)$, we found it convenient to focus on the annealed disorder averages as defined in \eqref{e:annealed}. These quantities have the advantage of changing sign at a certain value of the energy density $e$, which we define as the finite-$N$ `pseudo'-critical energies. These can be seen as finite-system counterparts of the critical energy in the thermodynamic limit. Although, for the small sizes studied, the actual values of the pseudo-critical energies are not close to the thermodynamic limit results, the trend is the correct one. An interesting outcome of these results is that the number of saddles is exponential in the system size for energies down to a critical value (N dependent) which is always larger than the ground state energy, i.e., this important property is not limited to the thermodynamic limit.

Strong finite size effects are also seen in the index-resolved cumulative complexities. Nevertheless, in agreement with the large-$N$ picture, the low energy sector is found to be dominated by small-index stationary points. In particular, at the lowest energies, only stationary points of index $I=0$ (i.e., minima) occur. Upon increasing the energy, stationary points with higher indices appear and gradually become more numerous than lower-index stationary points. This should also reflect in the relaxation properties of the model. In the large-$N$ limit, minima dominate completely for energy densities between the critical one $e_\text{c}$ and the threshold energy $e_\text{th}$. This leads to the well known two-step relaxation of the model in that energy regime. For small clusters we expect that a hierarchy of time scales will be present, which reflects the stability properties of the dominant set of stationary points at each energy.

Different from the quantities mentioned earlier, the mean index density shows rather weak finite-size effects. Its behavior is qualitatively similar to the large $N$ results, rising like a power law above the `pseudo'-threshold energy. From fits to our data we obtain an exponent $\alpha \approx 1.3$ for such a power law, reasonably close to the asymptotic result $\alpha=3/2$. From the same fits finite-$N$ approximations to the threshold energy can be extracted. These energies show the correct trend with increasing $N$ and are systematically larger than the corresponding pseudo-critical ones. The energy intervals so defined should be the small-$N$ analogs of the `minima-dominated' regime known to be present in the thermodynamic limit. If this has implications to the finite N dynamics is not known and is a point that certainly deserves to be studied.

Finally we studied the rescaled Hesse determinant $D$, evaluated at the stationary points of the 3SM. It is known, at least for certain models studied in the literature, that the vanishing of this quantity at some value of $e$ in the thermodynamic limit may signal the presence of a thermodynamic phase transition at that energy density. Despite the fact that the model shows a phase transition at $e_\text{c}$ in the thermodynamic limit, we found no evidence of a tendency to zero of the Hesse determinant at any value of $e$. There are many possible reasons for this to happen, the small system-sizes considered being one of them. Another reason may lie in the determinant criterion itself: It is based on the assumption that stationary points are 'at the origin' of a phase transition. Although this hypothesis is correct in many models, it is known to fail in others \cite{BaroniMSc,PhysRevE.70.036125,PhysRevE.70.041101,PhysRevLett.93.150601,PhysRevE.72.016122,PhysRevE.72.056134,PhysRevE.76.051119,Kastner:2011zz}, and in this case the reasoning behind the criterion breaks down.

The present paper focuses on properties of the potential energy landscape only. Since many of these properties are known to reflect also in dynamical and thermodynamical properties of finite systems, a study of these latter properties and a comparison to finite-$N$ energy landscape properties may lead to further insights on the interplay
between landscape and dynamics. This is a project worth being addressed in the future.

\acknowledgments
D.M.\ was supported by the U.S.\ Department of Energy under Contract No.\ DE-FG02-85ER40237, and he would like to thank Tarek Anous and Frederik Denef for introducing him to the $p$-spin model. The numerics were carried out on Fermilab USQCD cluster. M.K.\ acknowledges support by the Incentive Funding for Rated Researchers programme of the National Research Foundation of South Africa. D.A.S.\ acknowledges partial support from a research fellowship by CNPq, Brazil.

\bibliography{bibliography_NPHC_NAG}	

\end{document}